# Direct Detection of the Primordial Inflationary Gravitational Waves*

Wei-Tou Ni

Purple Mountain Observatory, Chinese Academy of Sciences

No. 2, Beijing W. Rd., Nanjing, 210008 China,

and

National Astronomical Observatories, Chinese Academy of Sciences

Beijing, 100012 China

e-mail: wtni@pmo.ac.cn

**SUMMARY**

Inflationary cosmology is successful in explaining a number of outstanding cosmological issues including the flatness, the horizon and the relic issues. More spectacular is the experimental confirmation of the structure as arose from the inflationary quantum fluctuations. However, the physics in the inflationary era is unclear. Polarization observations of Cosmic Microwave Background (CMB) missions may detect the tensor mode effects of inflationary gravitational waves (GWs) and give an energy scale of inflation. To probe the inflationary physics, direct observation of gravitational waves generated in the inflationary era is needed. In this essay, we advocate that the direct observation of these GWs with sensitivity down to $\Omega_{gw} \sim 10^{-23}$ is possible using present projected technology development if foreground could be separated.

*This essay received an "honorable mention" in the 2009 Essay Competition of the Gravity Research Foundation.*

## I. Introduction

Inflation is a rapid accelerated expansion which set the initial moments of the Big Bang Cosmology [1]. This expansion drives the universe towards a homogeneous and spatially flat geometry and accurately describes the average state of the universe. The quantum fluctuations in this era grow into the galaxies, clusters of galaxies and temperature anisotropies of the cosmic microwave background [2]. Although modern inflation originated from efforts of unification, its mechanism remains unclear. The quantum fluctuations in the spacetime geometry in the inflationary era generate GWs which would have imprinted tensor perturbations on the Microwave Background Radiation anisotropy. The analysis of five-year data of WMAP did not discover these tensor perturbations and showed that, combined with BAO and SN data, the tensor-to-scalar perturbation ratio r is less than 0.20 (95% CL) [3]. More than one order-of magnitude better accuracy in CMB polarization observation is expected from PLANCK mission [4] to be launched this year. Dedicated CMB polarization observers like B-Pol [5], EPIC [6] and LiteBIRD [7] missions would improve the sensitivity further by one order-of magnitude. This development would put the measurable r in the range of $10^{-3}$ corresponding to inflation energy scale of $10^{-16}$ GeV.

## II. Direct detection of inflationary GWs

If the tensor mode is detected in the CMB polarization observation, it could leads to two possible causes: (i) due to primordial gravitational waves [8]; (ii) due to (pseudo)scalar-photon interaction in the CMB propagation [9]. These possibilities may be distinguishable by detailed models or other implications [9]. Suppose we have possibility (i), more on inflationary physics can only be probed by direct gravitational-wave detection since next-generation CMB polarization observation will be limited by cosmic variances and cosmic shears [5-7].

For direct detection of primordial (inflationary, relic) GWs in space, one may go to frequencies lower or higher than the LISA [10] bandwidth, where there are potentially less foreground astrophysical sources to mask detection. DECIGO [11] and Big Bang Observer [12] look for GWs in the higher frequency range while ASTROD [13] and Super-ASTROD [14] look for GWs in the lower frequency range.

The intensity of a stochastic primordial background of GWs is usually characterized by the dimensionless quantity

$$\Omega_{GW}(f) = (1/\rho_c)\,(d\rho_{GW}/d\log f), \tag{1}$$

with $\rho_{GW}$ the energy density of the stochastic GW background and $\rho_c$ the present value of the critical density for closing the universe in general relativity. The minimum detectable intensity of a stochastic GW background $\Omega_{GW}^{min}(f)$ is proportional to detector noise power spectral density $S_n(f)$ times frequency to the third power [10, 15]. That is

$$h_0^2 \, \Omega_{GW}^{min}(f) \quad \sim \quad \text{const.} \times f^3 \, S_n(f), \tag{2}$$

where $h_0$ is the present Hubble constant $H_0$ divided by 100 km s$^{-1}$ Mpc$^{-1}$. Hence, with the same strain sensitivity, lower frequency detectors have an $f^{-3}$-advantage over the higher frequency detectors.

Figure 1 shows the estimated/predicted stochastic backgrounds with bounds and GW detector sensitivities. In the middle part of the diagram, the curves labeled 'Extragalactic' and 'Extrapolated' are the extragalactic GW foreground and extrapolated GW foreground from Farmer and Phinney [16]. The extension of WMAP constraint to higher frequency is according to a flat tensor index $r_t = 1$, and gives an upper bound at about $10^{-15}$ level. Dedicated future CMB polarization observers such as B-Pol mission [5], EPIC mission [6] and LiteBIRD [7] mission would improve the sensitivity to $10^{-16}$ level.

A typical strain sensitivity curve for a fixed period of integration for a space interferometric GW detector consists of three frequency regions; the antenna response region, the flat region, and the acceleration noise region. Toward high frequency region, once the armlength $L$ exceeds half the GW wavelength ($\lambda_{GW}$), the antenna response rolls down roughly as $\lambda_{GW}/L$, i.e., as $f^{-1}$. Hence, the sensitivity curve rises and the curve shifts to the left proportional to $L$. In the middle flat region, the sensitivity curve is dominated by the white photon shot noise. In the lower frequency region, the curve is dominated by the acceleration noise and it rises with some inverse power of frequency. As armlength $L$ increases, the sensitivity curve is shifted downward since the corresponding strain noise is inversely proportional to $L$. This is the main reason why ASTROD and Super-ASTROD can probe deep into inflation GW region.

A dedicated ASTROD GW detector has 3 spacecraft near Earth-Sun L3, L4 and L5 points forming a nearly equilateral triangle with armlength $2.6 \times 10^8$ km (52 times longer than LISA armlength). In figure 1, compared to LISA, ASTROD has 140,000 times ($52^3$) better sensitivity due to this reason. For a dedicated Super-ASTROD with 5 AU orbit, there is an additional 125 ($5^3$) times gain in sensitivity (the dotted line in between the ASTROD and Super-ASTROD sensitivity curves) compared to ASTROD due to longer armlength if the laser power and $S_n(f)$ stays the same. However, for baseline Super-ASTROD, the laser power is increased by 15 fold, and hence, the shot noise power spectral density is lowered by a factor of 15. Hence, the total gain in sensitivity due to $f^3 S_n(f)$ is a factor of 1875. In the study of GW background from cosmological compact binaries, Farmer and Phinney [16] showed that this background $\Omega_{gw}^{cb}(f)$ rolls off in the 1-100 μHz frequency region from $10^{-13}$ at 100 μHz to $10^{-17}$ level at 1 μHz. Therefore, Super-ASTROD will be able to detect this background and there is still ample room for detecting the primordial/inflationary GWs with optimistic amplitudes. For ASTROD, although the sensitivity curve extends well below the optimistic inflation line, the background from binaries [10] in the ASTROD bandwidth is as large as the most optimistic inflation signals. Signal separation methods should be pursued. If relic primordial/inflationary GW has a different spectrum than that cosmological compact binaries, their signals can be separated to certain degree. With this separation, Super-ASTROD will be able to probe deeper into the stochastic waves generated during inflation with less optimistic parameters. More theoretical investigations in this respect are needed.

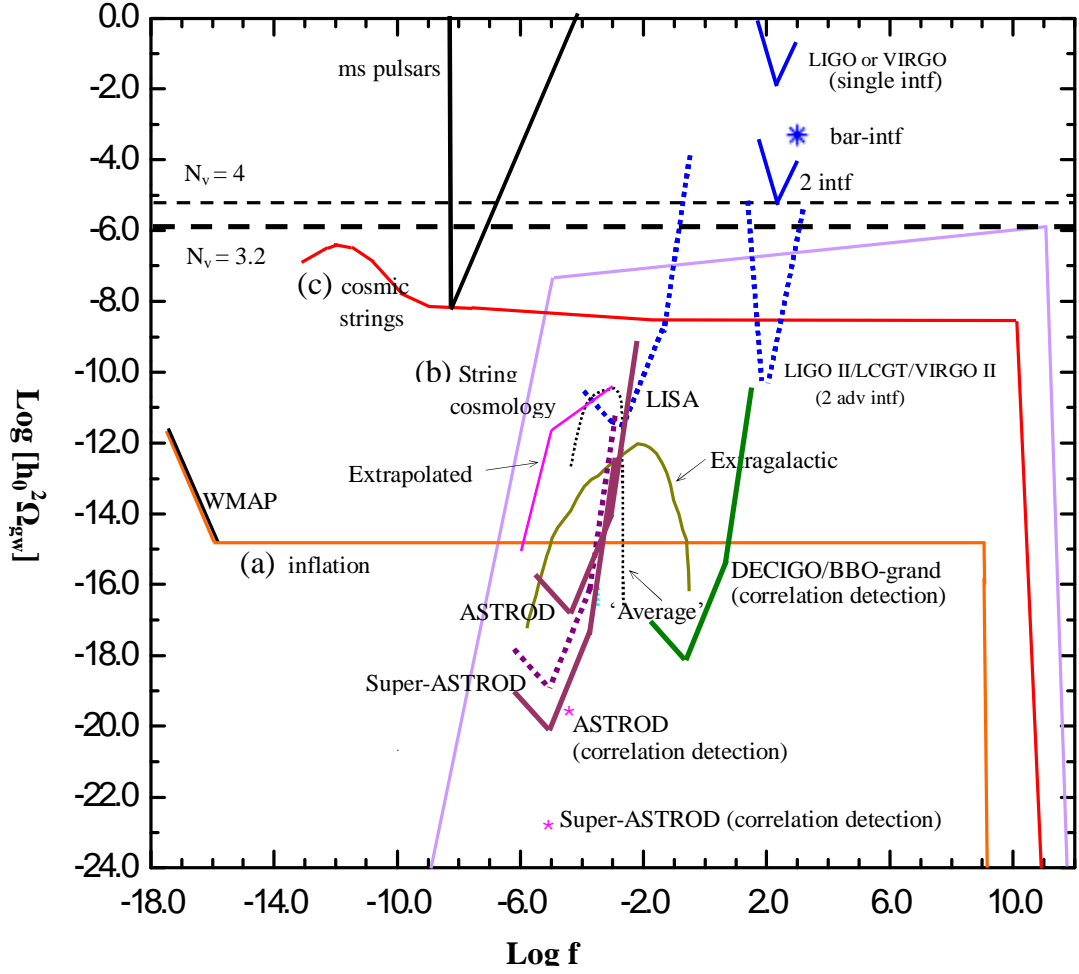

Figure 1. The stochastic backgrounds with bounds and GW detector sensitivities. (Adapted from Figure 3 and Figure 4 of [14, 15] with sensitivity curve of DECIGO/BBO-grand added; the extragalactic foreground and the extrapolated foreground curves are from [16]; see [14-16] and this section for explanations).

With correlation detection, DECIGO/BBO-grand propose to reach a sensitivity of $\Omega_{gw} \sim 10^{-18}$. With correlation detection, dual ASTROD and dual Super-ASTROD would reach a sensitivity of $\Omega_{gw} \sim 10^{-20}$ and $\Omega_{gw} \sim 10^{-23}$.

### III. OUTLOOK

Foreground separation and correlation detection method need to be investigated to achieve the sensitivities in figure 1. With current technology development, we are in a position to explore deeper into the origin of gravitation and our universe. The current and coming generations are holding such promises.